\documentclass[a4paper,aps,prl,showpacs,twocolumn,superscriptaddress]{revtex4}

\usepackage{graphicx,t1enc}
\usepackage{amsmath}

\begin{document}

\title{Dynamic Domain Networks}
\author{M. N. Kuperman}
\affiliation{Centro At{\'o}mico Bariloche and Instituto Balseiro,
8400 S. C. de Bariloche, Argentina} \affiliation{Consejo Nacional
de Investigaciones Cient{\'\i}ficas y T{\'e}cnicas, Argentina}
\author{M. Ballard}
\affiliation{Centro At{\'o}mico Bariloche and Instituto Balseiro,
8400 S. C. de Bariloche, Argentina}
\author{M. F. Laguna}
\affiliation{Centro At{\'o}mico Bariloche and Instituto Balseiro,
8400 S. C. de Bariloche, Argentina}

\begin{abstract}

We present a model for the description of the evolution of
contacts among individuals in a network. At each time step each
individual is associated with a domain or neighborhood of fully
connected agents.The dynamics of this changing neighborhood will
later be translated into a situation where the links between
individuals are also dynamic. A characterization in terms of the
parameters that govern the evolution of the network and a
comparison to previous work on Small World networks is presented
as well.
\end{abstract}

 \maketitle

\vspace{1cm}

\section{Introduction }

The study of networks and their applications within social
sciences has in recent years been a rich source of
interdisciplinary research. In particular, development of
mathematical models based on simple social interactions formulated
with graphs or networks, with certain topological properties, has
accumulated much interest. The concept of the small world was
first introduced by Milgram \cite{milgram} in an attempt to
describe the intrinsic properties of social communities and the
relationships that exist amongst the members thereof. A more mathematical
approach was later introduced to understand the underlying
aspects of Milgram's networks \cite{watts98}. Small World Networks
(SWN) are in essence, those networks which are comprised of a specified
amount of random and regular lattice connections. Driven by strong
evidence that certain complex networks share a close resemblance
with actual economic, social, and biological networks, a new field
has been developing within physics \cite{stro}. Such real world
applications have naturally led to the proposition of temporally
evolving networks \cite {ama,albe}. We will refer to
\textit{dynamic} networks as those which rewire links as time
progresses. Networks in which the links remain static in time once
the desired architecture has been achieved, are known as
\textit{quenched} networks. Whereas networks with links that
change randomly at every time step are called \textit{annealed}
networks. An example of the latter, which is akin to the model
developed here, may be found in \cite{manr}. The model we present
consists of subnetworks which are only linked to each other
through time. A dynamic domain network (DDN) is thus a type of
annealed network, though at a given moment the subnetworks or
domains are not connected to each other. As explained in the next
section, the model we propose may be associated with a phenomena
where the only mechanism of transmission or propagation throughout
the network is a type of diffusion. Though the model can be
generalized to include other features such as long range
transmission (as found in SWNs), we are for now interested in
analyzing properties due solely to this specialized diffusion. The
possibility of analyzing different dynamic degrees of freedom
independently will lead to a deeper understanding of diffusion
within networks. In the following sections we will consider in
more detail the features guiding our analysis,

\begin{itemize}

\item Two measurable parameters, $\omega$ and $\tau$, which may be associated with the SWN parameters, clusterization and
average path length.

\item Three types of domain movement, $p_d$, $p_f$, and $p_s$,
which are called the dynamic processes of the system, are analyzed

\item Numerical analysis and limited analytic calculation of
$\omega$ and $\tau$ as functions of the dynamic processes
demonstrate common characteristics with SWNs.

\item Diffusion coefficients associated with each dynamic process
are numerically and analytically calculated.

\item The DDN model reproduces features found in previous work on
disease propagation and further shows the affect of each dynamic
process on the total population infection.

\end{itemize}

\section{The model}
The model is based on a two dimensional array of individuals, each
one situated on the node of a matrix of size $n \times n$. The
system is  divided into $N$ domains, or fully connected
subnetworks, of size $k^2$. In what we call the regular case, all
domains have the same shape and size. We consider that an
individual is linked to the other individuals belonging to the
same domain, but not to the rest of the system. The boundaries
that demarcate the domains change in time and thus may be
considered as independent neighborhoods exchanging members through
dynamic links in time.  The architecture of the underlying dynamic
network is defined by the history of the changes of the
boundaries. A set of three processes govern the behavior of these
domains and thus three different parameters will be considered. At
each time step the whole set of domains may be displaced or
shifted from their former positions with probability $p_d$. It is
also possible that domains change individual size with probability
$p_s$ and change form or shape with probability $p_f$. Each of
these processes can act separately, but when all three take place
at the same time, a superposition of their effects occur. To
characterize some of the dynamic properties of the evolving
network we have chosen a few observable quantities as suggested in
\cite{manr}. We consider a measurable global variable analogous to
the average path length in SWNs. Over all domains and throughout
time, we analyze the time necessary for an `active' state to
propagate across the entire system. Initially, one individual is
chosen to be active. Any inactive individual in the same domain is
immediately activated by the presence of at least one active
individual in the same neighborhood. Once activated, the
individual remains as such for all time. We measure the time
involved $\tau_i$, to activate the node $i$ within the system. The
\textit{mean persistence} or \textit{activation time} $\tau$, is
calculated as defined in \cite{manr}. This value is studied for
different sets of $p_d$, $p_s$, and $p_f$. As a second
characterization of the system, we analyze the changes in the
composition of each individual's domain from one time step to
another. Again, in accordance with \cite{manr}, this local
measurable variable may be associated with a clustering
coefficient \cite{watts98}. As a general definition we call
$\omega_i(t)$ the \textit{overlap} or \textit{intersection} of
domains during $m$ consecutive time steps, where in the present
work $m=2$. The mean neighborhood overlap is then calculated for
varying values of $p_d$, $p_s$, and $p_f$. Compared to previous
models of dynamic networks, this model may be associated with
diffusive processes. This supposition will be reinforced with the
calculation of associated diffusion constants.

\section{Analytic considerations}
As a measure of the common area of two domains between consecutive
time steps, the overlap may be written analytically in a general
form. In the model we have considered that the allowable sizes
take on only quadratic values so that the square shape is always
accessible, particularly for the case when the form does not
change at all. We thus compare all possible forms of the permitted
sizes between a before domain and an after domain. The expression
for the mean overlap for any set of values
$\overline{p}=(p_d,p_s,p_f)$ can be written in general as a
binomial distribution
\begin{eqnarray}{\omega(\overline{p})}&=&\sum_{i=0}^2\,\sum_{j=0}^2\,\sum_{k=0}^2
\binom{2}{i} \binom{2}{j} \binom{2}{k} \label{overlap} \\
&& a_{ijk}(\overline{p}) p_d^i (1-p_d)^{2-i}p_s^j
(1-p_s)^{2-j}p_f^k (1-p_f)^{2-k} \nonumber
\end{eqnarray}
The coefficients $a_{ijk}(\overline{p})$ must be calculated by
considering the superposition for any succession of two allowed
states of the system for a given set of values $\overline{p}$.
Since we preserve the rectangular shape of the domains, the
overlap is determined by the divisors of domain sizes in the
before and after configurations. This allows us to write a
counting scheme for the overlap when $p_d = 0$.

\begin{equation}
    \frac{\sum_{m=\alpha}^{n}  \sum_{k=\alpha}^{n}  \left[  \sum_{i=\gamma_m}^{N_m}
    \sum_{j=\gamma_k}^{N_k}
     s_{ij}s_{(N_{m}-i+1)(N_{k}-j+1)}\frac{1}{ N_m N_k}\right]}{(n-\alpha+1)^2}.
\end{equation}
With this, the overlap is calculated for the four extreme
situations when $p_s$ and $p_f$ are either zero or one. The
parameters $N_m$ and $N_k$ indicate the \textit{number} of
divisors of the sizes $m^2$ and $k^2$, for each before and after
domain, respectively. The value $n^2$ is the maximum size allowed.
To calculate the case when $p_s=p_f=1$, one must set
$\alpha=\gamma_m=\gamma_k=1$; this sums over all forms and all
sizes bounded by $n$. For zero change in form and size
$p_s=p_f=0$,  the parameters must be set to $\alpha=n$,
$\gamma_m=\frac{N_m-1}{2}+1$, and $\gamma_k=\frac{N_k-1}{2}+1$.
Changes in size $p_s=1$, and no change in form $p_f=0$ correspond
to $\alpha=1$, $\gamma_m=\frac{N_m-1}{2}+1$, and
$\gamma_k=\frac{N_k-1}{2}+1$. Changes in form $p_f=1$, and no
change in size $p_s=0$, correspond to $\alpha=n$ and
$\gamma_m=\gamma_k=1$. These extreme values define four of the
eight coefficients in Eq. \ref{overlap}.

The underlying diffusive process occurring in the network as time
proceeds is another interesting feature we would like to address.
We begin by considering one particle located within an initial
domain. This test subject then performs a walk to neighboring
domains as the domain walls change and an overlap between the
original domain and a neighboring one permit movement of the
individual. Indeed, the process may be considered as a random
walk, characterized by a probability distribution particular to
each dynamic process. A brief analysis of what movement might be
expected can be done by considering the possibilities of change
associated with each process. As an example, consider the case of
a dynamic domain originally comprised of nine sites. A probability
can be calculated for a given random walker located at the center
of the original domain to step to the center of one of the
neighboring domains with which it overlaps in a given time step.
Two different jumps should be considered; those to neighboring
domains that share a side with the original one, and those to any
of the other four domains along the diagonals. We will call
$\alpha$ the probability of the first type of jump, and $\beta$
the probability corresponding to jumps of the second type. The
$\alpha$ and $\beta$ probabilities for each of the three processes
can be calculated by considering the fractions of all the overlap
situations that will lead to a given jump, where we have
calculated $\alpha_d=0.12$, $\beta_d=0.04$, $\alpha_s=0.024$,
$\beta_s=1/256$, $\alpha_f=1/9$, $\beta_f=0$. These values are
used to evaluate a random walk process and thus an associated
diffusion coefficient.

A random walker in one dimension with equal likeliness to jump to
the right or left is analyzed. This probability to move is given
by $p(\alpha + 2 \beta)$, where $\alpha$ and $\beta$ correspond to
the values mentioned previously and $p$ is the value
characterizing the probability for a change in the position, size,
or shape. Further, there is a probability to stay in the same
site, $\gamma=1-p(2\alpha+4\beta)$. This leads to a calculation of
the mean square displacement $\langle x^2 \rangle$, of a set of
particles equal to $18 p(\alpha+2 \beta) t$ \cite{manuel}. By
considering the appropriate values for each dynamic process, we
find that the diffusion coefficients for each are given by
$D_d=1.8 p_d$, $D_s=0.28 p_s$ and $D_f=p_f$.

\section{Numerical results}

\begin{figure}[hbt]
\centering \resizebox{\columnwidth}{!} {\rotatebox[origin=c]{0}{
\includegraphics{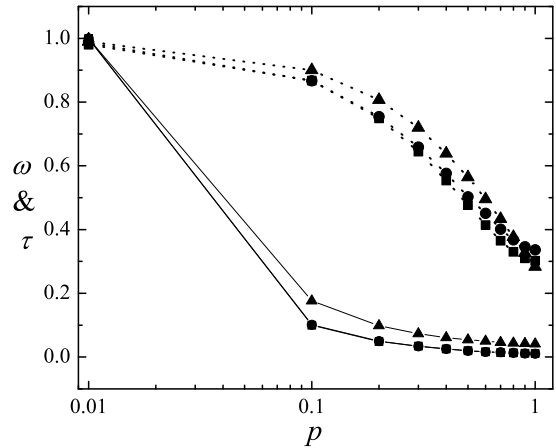}}}
\caption{ The bold line ${\tau}$, and the dotted line ${\omega}$
as functions of $p$, where $p=p_d$(squares), $p=p_s$ (triangles),
and $p=p_f$ (circles). Non-varying $p$ values are set equal to
zero. ${\tau}$ is normalized to the total activation time when
$p=10^-2$.} \label{onep}
\end{figure}

Extensive numerical simulations of the
described model for networks of $n^2=10^4$ to $10^7$ nodes and $k=3$ to $10$ have been
performed. A typical realization starts with the random election of an active individual.
In the successive time steps, the active condition propagates throughout the system until
the system is entirely activated. We calculate the mean activation time of the system as
$${\tau}= \frac{1}{n^2}\sum_{i=1}^{n^2} \tau_i,$$ where $\tau_i$ is the activation
time of the $i$ node. We also measure the mean neighborhood
overlap between two consecutive time steps as
$${\omega}= \frac{1}{\tau_A n^2} \sum_t^{\tau_A} \sum_i^{n^2} \omega^{(2)}_i(t),$$ where the sum over time
is performed over all the time steps until total activation at
time $\tau_A$. In Fig.\ref{onep} we plot the values of ${\tau}$
and ${\omega}$ for the specific cases when two of the three $p$
values are zero and the remaining one varies.

\begin{figure}[hbt]
\centering \resizebox{\columnwidth}{!} {\rotatebox[origin=c]{0}{
\includegraphics{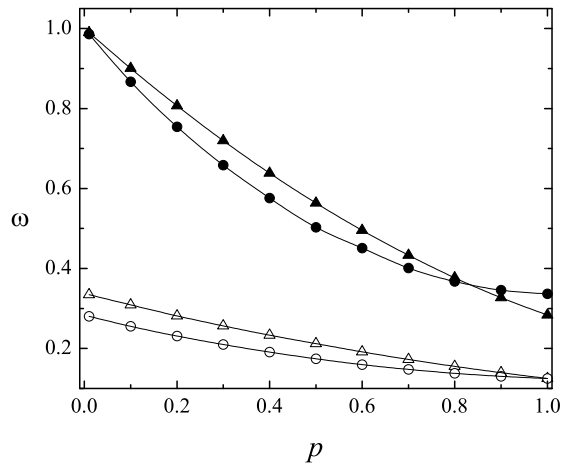}}}
\caption{The average overlap ${\omega}$, as a function of $p$,
where $p=p_s$ (triangles), $p=p_f$ (circles), and $p_d=0$ in all
cases plotted here. Solid symbols indicate that the non-varying
$p$ parameter is one, whereas open symbols indicate the
non-varying $p$ parameter is zero.} \label{numan}
\end{figure}

A comparison between the analytic calculations of ${\omega}$ and
the numerical results are displayed in Fig. \ref{numan}. Again,
only the limiting cases are considered. The lower curves
correspond to changing either $p_s$ or $p_f$, while $p_d$ and the
remaining $p$ value are both zero. Correspondingly, the upper
curves depict a varying $p_s$ or $p_f$, while $p_d$ is again equal
to zero, but the remaining term is now constantly one. The
continuous line is obtained by analytic calculation on top of
which we have plotted the numerical results.

\begin{figure}[hbt]
\centering \resizebox{\columnwidth}{!} {\rotatebox[origin=c]{0}{
\includegraphics{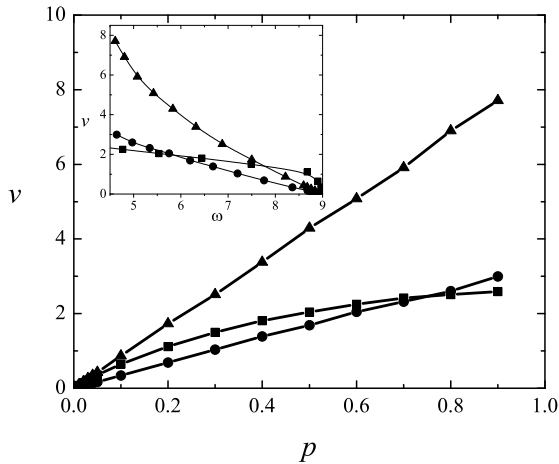}}}
\caption{The average velocity of propagation $v$, as a function of
$p$, where $p=p_d$(squares), $p=p_s$ (triangles), and $p=p_f$
(circles). In the inset we plot $v$ as a function of ${\omega}$}
\label{velo}
\end{figure}

An interesting aspect that leads us to consider diffusion within the system,
is the evolution of the size of the cluster of active
individuals. For this we have performed several calculations considering an initial active
agent located at the center of the system. For several
values of $\overline{p}$ and considering the variance of only one $p$
parameter at a time, we have evaluated the mean radius of the growing
nucleus. The radius evolves linearly with time and thus
the velocity of propagation of the active state $v$, is easily calculated. In Fig. \ref{velo}, $v$ is depicted as a function of $p$ for each dynamic process. The
inset shows $v$ as a function of the mean overlap. Apparent from the figure, the velocity adopts
a different behavior for each of the three dynamic processes.

\begin{figure}[hbt]
\centering \resizebox{\columnwidth}{!} {\rotatebox[origin=c]{0}{
\includegraphics{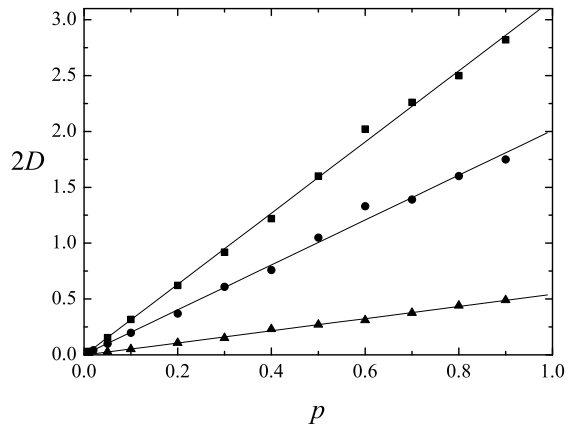}}}
\caption{Diffusion coefficients as functions of $p$, where
$p=p_d$(squares), $p=p_s$ (triangles), and $p=p_f$ (circles).
Solid lines indicate the analytic calculation over which the
numerical results are plotted.} \label{difu}
\end{figure}

In conjunction with analytic calculations of the diffusive process, we have performed
numerical simulations to measure the mean
square displacement $\langle x^2 \rangle $, where $\langle$
$\rangle$ indicates the mean value over individuals and $x$ is the
position of a random walker on the $x$ axis. From the analysis of
this quantity throughout time, we obtain a value for the diffusion
coefficient. Fig. \ref{difu} shows this value as a function of
each of the three processes defined by $p_s$, $p_d$, and $p_f$.
To compare numerical results with the $\alpha$ and $\beta$ values
obtained before, we have considered several individuals
performing a random walk in the system. Walkers move along the
centers of the domains. A given jump is allowed only if
superposition between two domains occurs in two successive time
steps. The results correspond to networks undergoing the different
processes governed by the aforementioned $p$ parameters.
The fitting by minimum squares methods gives us the following
values: $D_d=1.6 p_d$, $D_s=0.27 p_s$ and $D_f=p_f$, which are in
agreement with previously calculated values.

\section{Disease propagation }

\begin{figure}[hbt]
\centering \resizebox{\columnwidth}{!} {\rotatebox[origin=c]{0}{
\includegraphics{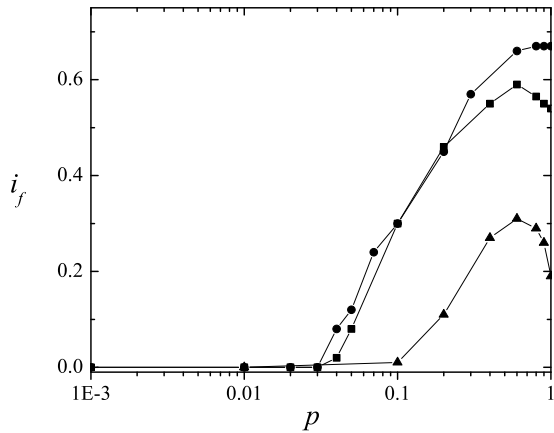}}}
\caption{ The fraction of infected individuals with respect to the total population
$i_f$, as a function of $p$, with $p=p_d$(squares), $p=p_s$ (triangles), and $p=p_f$
(circles)} \label{inf1}
\end{figure}

The DDN model that we have proposed compares to previously
obtained results of disease propagation within Dynamic Small World
Networks \cite{zan} in the following way. We consider a standard
model \cite{Murray} for an infectious disease with three stages:
susceptible (S), infectious (I), and refractory (R). Any
susceptible individual can become infected with a given
probability by an infected individual within the same domain or
neighborhood. The infection cycle ends when the element reaches
the refractory state after $\tau$ time steps. The refractory state
is a permanent condition and thus the individual cannot be
infected again. The algorithm goes as follows. At each step an
infected element $i$ is chosen at random. If the time elapsed from
the moment $t_i$ when it entered the infection cycle up to the
current time $t$ is larger than the infection time $\tau$, the
element $i$ becomes refractory. Otherwise, one of its neighbors
$j$ of $i$ is randomly selected. If $j$ is in the susceptible
state, contagion occurs. Element $j$ becomes infected and its
infection time $t_j=t$ is recorded. If on the other hand, $j$ is
already infected or refractory, it preserves its state. Since each
time step corresponds to the choice of an infected individual, the
update of the time variable depends on the number $N_I(t)$ of
infected individuals at each step, $t\to t + 1/N_I(t)$. Once there
is no change in the number of infected individuals for the
duration of time necessary for all infected individuals to
transform into the refractory state $\tau$, then activation
time has been achieved and the process stops. Here we set $\tau=3$
for the infection time which insures that for intermediate values
of the $\overline{p}$ parameters, the disease spreads over a
finite fraction of the population. Moreover, we consider an
initial condition where there is only one infected element; all
the other elements being susceptible. The considered initial
condition thus represents an initially localized disease. The
infection will remain localized during the initial stages and will
propagate according to the behavior of the domains. In Fig.
\ref{inf1} we plot the proportion of total infected individuals
$i_f$ as it varies with different values of $p$. Two of the $p$ values are maintained fixed while the remaining one
varies between zero and one.

In comparison with the results obtained in \cite{zan}, we find
that the threshold values of $p$ for disease propagation are of
the same order.  The sharp transition from the non-propagative
regime to the propagative one as a function of the structure of
the network is one of the most interesting aspects presented by
epidemiological models on networks. The maximum proportion of
infected individuals shows that a fraction of individuals remain
not infected. Similar values of final infected individuals have
been obtained in \cite{zan}. The non-monotonic behavior of this
quantity has also been observed in other epidemiological models
based on Small World Networks \cite{kup}. A possible explanation for
the decrease of infection for large values of $p_s$ involves the
fact that in the DDN system, there is a bias towards decreasing
the size of a domain. A high level of dynamics will temporally
isolate infected individuals. It is because of this limitation of
the model that the infection cannot propagate as efficiently for
large $p_s$ and a smaller number of infected individuals is detected.

\section{Conclusions}

If we aim to compare the system presented here with previous work
on dynamic Small Worlds, we should limit this comparison to cases
when shortcuts only to relatively close neighbors are considered.
This restriction is due to the fact that in the present model, new
links can only be established with nodes occupying neighbor
domains. An interesting aspect despite these limitations is that
we still observe some of the features found in dynamic SWN models
without such restrictions. In the first part we have studied the
behavior of the mean persistence time ${\tau}$, which is similar
to the SWN path length, and mean neighborhood overlap ${\omega}$,
analogous to the SWN clusterization. Though no analysis of how the
domain size scales with these quantities was done, the qualitative
similarities ought to be noted. That is, there exists a region in
the $p$ parameter space when the system displays a mixture of the
two limiting cases, relatively high overlap and relatively low
activation time. The value of ${\omega}$ can be analytically
obtained for some limiting cases. Further, we have studied the
underlying diffusive process occurring on the network. A diffusion
coefficient was numerically as well as analytically calculated for
the three dynamic processes, $p_s$, $p_f$, and $p_d$ and
demonstrated the diffusive contribution of each. As a final
application and point of comparison with previous work, we analyze
the propagation of an infection. For small $p$ values, an initial
infection can not propagate. At a given $p$ value, propagation is
possible and grows rapidly to a maximum infected fraction of the
entire population as $p$ increases. The threshold values are
similar to those previously observed for a dynamic SWN. In this
case, we show that despite the fact that there is a transition
around the same value of $p$, the behavior of the system strongly
depends on which dynamic domain process governs the dynamics of
the system.

\section{Acknowledgements}

M.B. would like to thank the OAS for partial support.

\end{document}